\documentclass[prl,aps,showpacs,superscriptaddress,twocolumn]{revtex4}
\usepackage{graphicx}
\usepackage{epsfig}
\usepackage{amssymb}
\usepackage{amsmath}
\usepackage{amsfonts}
\usepackage{color}

\begin{document}
\title{Generation of mesoscopic entangled states in a cavity coupled to an atomic ensemble}
\author{G. Nikoghosyan}
\affiliation{Institut f\"ur Theoretische Physik, Albert-Einstein Allee 11, Universit\"at
Ulm, 89069 Ulm, Germany}
\affiliation{Institute of Physical Research, 378410, Ashtarak-2, Armenia}
\author{M.J. Hartmann}
\affiliation{Technische Universit\"at  M\"unchen, Physik Department, James-Franck-Strasse, 85748 Garching, Germany}
\author{M. B. Plenio}
\affiliation{Institut f\"ur Theoretische Physik, Albert-Einstein Allee 11, Universit\"at
Ulm, 89069 Ulm, Germany}
\date{\today }

\begin{abstract}
We propose a novel scheme for the efficient production of "NOON states" based on
the resonant interaction of a pair of quantized cavity modes with an
ensemble of atoms. We show that in the strong-coupling regime the
adiabatic evolution of the system tends to a limiting state that describes mesoscopic entanglement
between photons and atoms which can easily be converted to a purely
photonic or atomic NOON state. We also demonstrate the remarkable property that the efficiency of this
scheme increases exponentially with the cavity cooperativity factor, which gives efficient access to high number NOON states. The experimental feasibility of the scheme is
discussed and its efficiency is demonstrated numerically.
\end{abstract}

\pacs{42.50.Gy, 42.50.Pq, 42.65.Lm, 03.67.Bg}
\maketitle



There is considerable interest in the development of mesoscopic entangled states
as resources for quantum technology applications. Of particular interest are entangled
states for $N$ photons of the form $\left\vert N\right\rangle _{a}\left\vert
0\right\rangle _{b}+\left\vert 0\right\rangle _{a}\left\vert N\right\rangle
_{b}$ (NOON states), which contain $N$ indistuinguishable particles in an
equal superposition of all being in one of two possible modes $a$ and $b$.
A number of applications for these states have been suggested, including
entanglement enhanced metrology and sub-wavelength lithography
\cite{Edamatsu-PRL-2002, Fonseca-PRL-1999, Wineland-PRA-1996, Boto-PRL-2000, Shih-PRL-2001, Huelga-PRL-1997}.
Yet building sources for such states is challenging because their
decoherence rate increases linearly with $N$.

Recently NOON states with up to $N=3$ microwave photons have been generated deterministically
\cite{Yamamoto-2011}. In the optical domain however only few experimental realizations were
reported to have generated $N>2$ NOON states, which were mostly based on linear optics
and state projective measurements of photons which implies an exponential scaling
of resources with $N$ \cite{Pryde-PRA-2003, Steinberg-Nature-2004}. For some other schemes
perfect optical elements and good initial sources of quantum fields are required
\cite{Dowling-PRL-1993, Kwiat-PRL-2009, Fiurasek-PRL-2002, Takeuchi-Science-2007, Silberberg-Science-2010}.

In this letter we propose a scheme which leads to a robust unitary evolution
into highly nonclassical entangled states of atoms and photons. These states can then be
converted into a photonic NOON state by application of a single projective
measurement on the atoms which can be implemented more efficiently than projective measurements
on photons. Moreover the transit time of our system between NOON states with $N$ and $N+1$
excitations decreases linearly in $N$. This can be used to compensate the
increasing decay and dephasing rate of NOON states and thus to generate high photon number NOON states
efficiently and on demand.

We start by introducing the physical setup and the basic idea of the mechanism that
we intend to exploit. Then we present analytical explanations and estimates for the
feasibility range of our scheme, which are backed-up by a careful numerical study that
confirms our findings. We finish off with conclusions and an outlook.

\begin{figure}[b]
\begin{center}
\includegraphics[width=9cm]{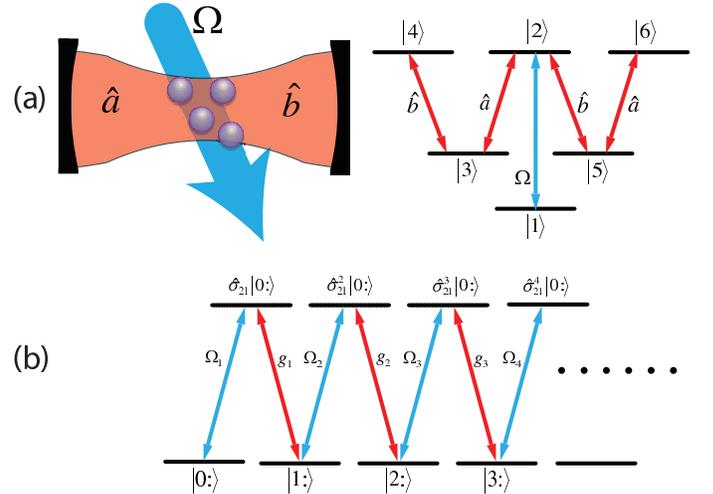}
\end{center}
\caption{(Color online) (a): Arrangement of atoms in the cavity and atomic level structure. (b): Schematic diagram of coupling of relevant $|n:\rangle$ states (see the definition in the text).
Here, $\Omega_n = \Omega \sqrt{M-n+1}$ and $g_n = g \sqrt{n(n+1)}$.
The initial state $\left\vert \psi
_{in}\right\rangle$ is identical to $|0:\rangle$. The dark state (\ref{dark_state}) is a coherent superposition of $|n:\rangle$ states.
Due to the destructive interference the population of excited states is negligible.}
\label{fig:1}
\end{figure}

{\em Hardware: The physical setup --} We consider a cloud of atoms with the
level structure depicted in Fig.\ref{fig:1}a that interact with a classical pump
field and a pair of quantized cavity modes (Fig.\ref{fig:1}a). The pump field that
runs at some angle
to the cavity axis resonantly couples to the transition $|1\rangle
\leftrightarrow |2\rangle$. The transitions $|2\rangle \leftrightarrow |3\rangle$
and $|5\rangle \leftrightarrow |6\rangle$ ($|2\rangle \leftrightarrow |5\rangle $
and $|3\rangle \leftrightarrow |4\rangle $) are driven by cavity mode $a$ ($b$).
This setup can for example be realized with a cavity tuned to the D1 line
of $^{87}$Rb atoms. Then upper states $|2\rangle$, $|4\rangle$, $|6\rangle$ are, respectively the Zeeman $m=0, -2, +2$ sub-levels of
hyperfine states with total angular momentum $F'=2$, lower states $|3\rangle$, $|5\rangle$ correspond to $m=-1, +1$ of $F=2$ sublevels and $|0\rangle$ can for example be $|0\rangle$ of $F=1$. A suitable choice
for the two cavity modes are thus the two orthogonal polarizations $\sigma^{+}$
and $\sigma^{-}$ of the same frequency that are supported by the cavity. Other Zeeman sub-levels can be neglected since they are not resonantly coupled to the considered states.
Such systems are routinely set up with current technology and are thus available in
several laboratories
\cite{cavity-QED-Kimble, Esslinger-Nature-2007, Stamper-Kurn-2007, Kuhn-PRL-2000, Vuletic-PRL-2010, Rempe-PRL-2009, Vuletic-Science-2011, Meschede-PRL-2009, Reichel-Nature-2007}.

{\em Physical principles and analytical estimates --}
We will now turn to the mathematical description of the physical
mechanisms that are underlying our scheme. To simplify the discussion
the spatial structure of the cavity mode as well as the coupling strength
for all transitions is assumed to be the same. We also assume that the
interaction time is much shorter than the lifetime of a photon in the
cavity mode. In this case the Hamiltonian of our system reads in a
suitable rotating frame,
\begin{equation}
\hat{H}=\hbar \left[ \Omega \left( t\right) \hat\sigma _{21}+g\hat{a}\left( \hat{%
\sigma}_{23}+\hat{\sigma}_{65}\right) +g\hat{b}\left( \hat{\sigma}_{25}+\hat{%
\sigma}_{43}\right) +H.c\right]  \label{Hamiltonian}
\end{equation}
where, $\Omega \left( t\right) $ is the Rabi frequency of the pumping, $\hat{a}$
and $\hat{b}$ denote the annihilation operators for the cavity modes,
$g$ is the vacuum Rabi frequency and $\hat{\sigma}_{mn}=\sum\limits_{i=1}^{M}\left\vert m\right\rangle_{ii} \left\langle n\right\vert $ are the collective atomic flip operators for $M$ atoms.

The Hamiltonian (\ref{Hamiltonian}) has a dark state, i.e. an eigenstate which
does not contain the short-lived upper states $|2\rangle, |4\rangle$ and
$|6\rangle$,
\begin{equation}
\left\vert D\right\rangle= \frac{1}{\sqrt{C}} \sum\limits_{n=0}^{M-1}\frac{1}{\sqrt{n!(n+1)!}}%
\left( -\frac{\Omega \left( t\right) }{g}\right) ^{n}\left\vert
n:\right\rangle  \label{dark_state}
\end{equation}
where $C = \sum\limits_{n=0}^{M-1}\frac{1}{n!(n+1)!}\left( -\frac{\Omega \left( t\right) }{g}\right) ^{2n}$ is a normalization constant and
we have introduced the notation
\begin{equation}
|n:\rangle=\frac{1}{\sqrt{n!(n+1)!}}\left\{\alpha ( \hat{a}%
^{\dagger})^n \hat{\sigma}_{31}^{n+1} + \beta (\hat{b}^\dagger)^{n} \hat{\sigma}_{51}^{n+1}\right\} |vac\rangle  \label{k_state}
\end{equation}
where $\left\vert vac\right\rangle $ is the state where all atoms of the system
are in level $\left\vert 1\right\rangle $ and there are no photons in the cavity.
The states (\ref{k_state}) are given by a mesoscopic
superposition of atoms in state $\left\vert 3\right\rangle $ and photons in
mode $\hat{a}$ superimposed with atoms in state $\left\vert 5\right\rangle $
and photons in mode $\hat{b}$. The dark state nature of eq. (\ref{dark_state})
emerges from two arguments. The excited states $|4\rangle (|6\rangle)$ of the
atoms remain empty as one would need to start from a state with atomic excitations
in level $|3\rangle$ and photons in mode $\hat{b}$ (atomic excitations in level
$|5\rangle$ and photons in mode $\hat{a}$) to excite them. Neither of these
configurations feature in the dark state eq. (\ref{dark_state}). Excitations of
the atomic level $|2\rangle$ in turn can not be created since the two pathways
$|n:\rangle \leftrightarrow \hat{\sigma}_{12} |n:\rangle$ and  $|n+1:\rangle \leftrightarrow
\hat{\sigma}_{12} |n:\rangle$ for exciting a state $\hat{\sigma}_{12} |n:\rangle$
interfere destructively \cite{Harris-PRL-1989}, see Fig.1b. For these reasons,
(\ref{dark_state}) is indeed a dark state \cite{Yudin-EPL-2005}.

The basic physical mechanism that is proposed here also assumes that
our system is initially prepared in the state $\left\vert \psi
_{in}\right\rangle =\left( \alpha \hat{\sigma}_{31}+\beta \hat{\sigma}
_{51}\right) \left\vert vac\right\rangle $ $\left( \left\vert \alpha
\right\vert ^{2}+\left\vert \beta \right\vert ^{2}=1\right) $.
This initial state can be, for instance, prepared by application of a weak pump
field via a cavity mediated adiabatic passage process \cite{Rempe-PRL-2009}.
For $\Omega =0$, the dark state eq.(\ref{dark_state}) is identical to the
initial state $\left\vert \psi _{in}\right\rangle $ whereas population is transferred to states $|n:\rangle$ with increasing intensity of the
pumping field. Thus if the pump field is switched on adiabatically the system
remains in the dark-state $|D\rangle$. Initially, when the cavity is empty the
adiabaticity condition reads,
\begin{equation}
    \frac{g^{2}}{\Gamma }\gg \left\vert \frac{\dot{\Omega}( t) }
    {\Omega(t)}\right\vert  \label{adiabatic}
\end{equation}%
where $\Gamma $ is the relaxation rate of upper states, which for the sake of
simplicity is assumed to be equal for all atomic transitions. However, when the
system contains more than $1$ photon the effective coupling between the atoms
and the photons is increased by a factor of $\sqrt{n_{ph}}$ and one expects
the adiabaticity condition to generalize to
\begin{equation}
\frac{g^{2}n_{ph}}{\Gamma }\gg \left\vert \frac{\dot{\Omega}( t) }
{\Omega} \right\vert.  \label{adiabatic2}
\end{equation}%
While we have not proven this condition analytically due to the complex
energy level structure, it is in excellent agreement with our numerical results.

From eq.(\ref{dark_state}) one observes that the population of states with $n_{ph}$
excitations depends on the amplitude of the pumping field. Moreover when the
pump field becomes larger than $g$ the main part of the population
accumulates in states with
\begin{equation}
    n_{ph}\approx \frac{\Omega }{g}  \label{photon_number}
\end{equation}%
Thus the adiabaticity condition (\ref{adiabatic2}) is easier to satisfy when
the amplitude of the pumping field is larger. From eq.(\ref{dark_state}) one
can also see that if $\Omega \gg gM$ all the population is transferred to the states
with the highest possible number of excitations $\left( n=M-1\right) $. Then
our dark state is reduced to
\begin{equation}
    |D\rangle = |(M-1):\rangle.  \label{dark_state2}
\end{equation}
which is a maximally entangled state with $M$ excitations either in the
left leg ($\hat{a}$-photons and atoms in level $|3\rangle$) or the
right leg ($\hat{b}$-photons and atoms in level $|5\rangle$) of the system.

Although this state is an entangled state of atoms and photons it can
easily be converted to purely photonic (atomic) NOON states by projective
measurements and local unitary operations. For example resonant quick
$\pi /2$ microwave pulses between Zeeman sub-levels $|3\rangle$ and $|5\rangle$
can be used to map $\hat{\sigma}_{31} \to \frac{1}{\sqrt{2}}(\hat{\sigma}_{31}+\hat{\sigma}_{51})$
and $\hat{\sigma}_{51} \to \frac{1}{\sqrt{2}}(\hat{\sigma}_{31}-\hat{\sigma}_{51})$.
This implies $ |D\rangle \to
\left\{\alpha ( \hat{a}^{\dagger})^{M-1} (\hat{\sigma}_{31}+\hat{\sigma}_{51})^{M} + \beta (\hat{b}^\dagger)^{M-1} (\hat{\sigma}_{31}-\hat{\sigma}_{51})^{M}\right\} |vac\rangle
= \left[\alpha ( \hat{a}^{\dagger})^{M-1} + \beta (\hat{b}^\dagger)^{M-1} \right] \sum\limits_{k=0}^{M/2}\binom{{M}}{{2k}}\hat{\sigma}_{31}^{M-2k}\hat{\sigma}_{51}^{2k}|vac\rangle +$
$\left[\alpha ( \hat{a}^{\dagger})^{M-1} - \beta (\hat{b}^\dagger)^{M-1} \right] \sum\limits_{k=1}^{M/2}\binom{{M}}{{2k-1}}\hat{\sigma}_{31}^{M-2k+1}\hat{\sigma}_{51}^{2k-1} |vac\rangle
$
where we have skipped the normalization for simplicity. Hence the detection
of $K$ quanta in level $|3\rangle$ projects the photons into the pure
photonic NOON state.
\begin{equation*}
    |NOON\rangle = \frac{1}{\sqrt{\alpha^2 + \beta^2}}\left[\alpha(a^{\dagger})^{M-1} + (-1)^{K} \beta(b^{\dagger})^{M-1}\right]|vac\rangle
\end{equation*}

Making use of the cyclic transitions $|3\rangle \leftrightarrow |5\rangle$ the projective
measurement can be done with high fidelity. If the initial number of atoms is known the detection
of $K$ atoms in level $|3\rangle$ produces a NOON state with probability proportional to
$p^K\sum\limits_{m=K/2}^{(M-K)/2}(1-p)^{2m}\binom{{M}}{{2m}}$ where, $p$ is the
probability for the successful detection of an excitation in level $|3\rangle$. It should be
noted here that the efficiency of atomic state detection ($p>0.9$) is much higher than the
typical efficiency of single-photon detectors ($p\sim0.4$). Hence for the preparation of high NOON states, which on average require the detection of several quanta in state $|3\rangle$, our scheme performs dramatically better than schemes that rely on the detection of multiple individual photons.

{\em Estimate of limitations --}
Important practical limitations of the present scheme result from
dissipation in the form of cavity damping and spontaneous emission. Cavity
damping comes into play as soon as the cavity mode is excited and causes
decoherence. It can be neglected if
\begin{equation}
    \kappa \int_{0}^{T}n_{ph}\left( t\right) \mathrm{d}t\ll 1
    \label{cavity_loss}
\end{equation}%
where, $\kappa $ is the cavity decay rate, $T$ is the duration of the
process and $n_{ph}\left( t\right) $ is the number of photons in the cavity
at time $t$. Spontaneous decay can be disregarded if the
interaction is adiabatic i.e. satisfying (\ref{adiabatic2}). Thus the scheme works if
both conditions (\ref{adiabatic2}) and (\ref{cavity_loss}) are satisfied.
Integrating both sides of inequality (\ref{adiabatic2}) yields
$\frac{g^2}{\Gamma} \int_{t_1}^{T}n_{ph}\left( t\right) \mathrm{d}t \gg
\ln\left( \frac{\Omega(T)}{g} \right) $
where $t_1$ is the time when the cavity contains one photon ($\Omega(t_1)= g$).
Then by making use of relations (\ref{photon_number}) and (\ref{cavity_loss}) one
can show that the scheme is expected to work if
\begin{equation}
    \frac{g^{2}}{\kappa \Gamma }\gg \ln M.  \label{cavity_decay}
\end{equation}
The term on the left hand side is the so called cooperativity parameter which
represents the coupling strength between single atom and a cavity photon. Thus
in the presented system the size of the achievable NOON state increases
exponentially for increasing cooperativity. State-of-the-art technology enables
production of cavities with cooperativity of order of 10, which according to
condition (\ref{cavity_decay}) will allow efficient generation of NOON states
with $n_{ph} \lesssim 10$.

{\em Numerical verification --}
We will now present a numerical analysis of the system to assess its feasibility in detail. For an initial state of the form  $\left\vert \psi_{in}\right\rangle =\left( \alpha \hat{\sigma}_{31}+\beta \hat{\sigma}_{51}\right) \left\vert vac\right\rangle $,
all relevant states of our system can be written as
\begin{align}
& |\psi \rangle
=\sum\limits_{m=0}^{M}\sum\limits_{l=0}^{M-m}\sum\limits_{k=0}^{M-m-l}\sum%
\limits_{q,r=0}^{\min \left\{ k,l\right\} }[\alpha A_{m,k,l,q,r}\left(
t\right) \hat{\sigma}_{31}+  \notag \\
& +\beta B_{m,k,l,q,r}\left( t\right) \hat{\sigma}_{51}]\left\vert
m,k,l,q,r\right\rangle  \notag
\end{align}%
where $A_{m,k,l,q,r}\left( t\right) $ and $B_{m,k,l,q,r}\left( t\right) $
are the time dependent probability amplitudes and
$\left\vert m,k,l,q,r\right\rangle =\hat{\sigma}_{21}^{m}\hat{\sigma}_{31}^{k-q}\hat{\sigma}%
_{51}^{l-r}\hat{\sigma}_{41}^{q}\hat{\sigma}_{61}^{r}\left( \hat{a}^{\dagger }\right)
^{k-r}\left( \hat{b}^{\dagger }\right) ^{l-q}\left\vert vac\right\rangle
$
The evolution of the system
is described by the time dependent Schr\"{o}dinger equation with Hamiltonian
(\ref{Hamiltonian}). To include the
decay of excited states $\left\vert 2\right\rangle $, $\left\vert
4\right\rangle$, and $\left\vert 6\right\rangle $ we assume that the
amplitudes containing these states decay exponentially with rate
$(m+q+r)\Gamma$, where the term in the brackets is the number of atoms in
the excited states.

\begin{figure}[tbp]
\begin{center}

\includegraphics[width=9cm]{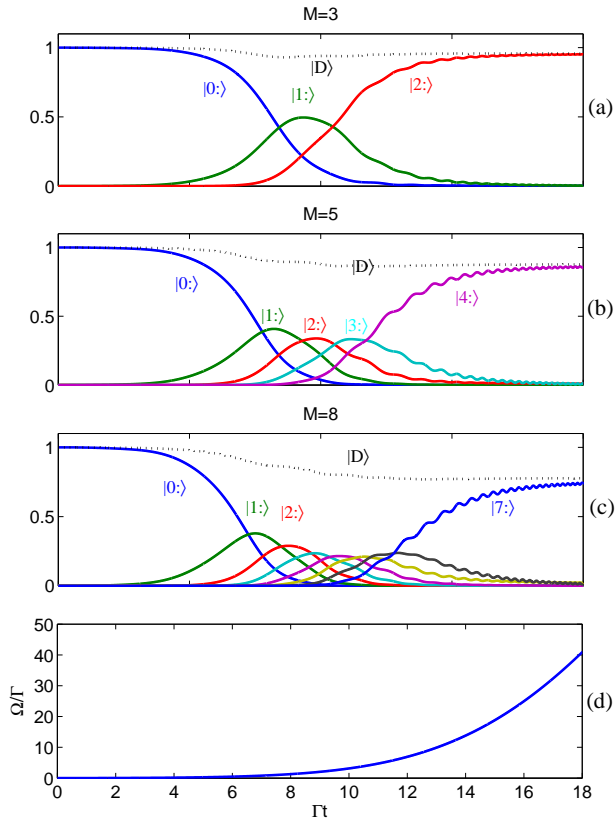}
\end{center}
\caption{(Color online) Time evolution of the system: For 3 (a), 5 (b),
and 8 (c) atoms. The dashed black curve is the population in dark-state. The
pumping field has been chosen $\Omega /\Gamma =10^{3} \exp\left[ -0.01(\Gamma
t-30)^{2}\right] $ its leading edge is presented in (d). }
\label{fig:2}
\end{figure}
\begin{figure}[tbp]
\begin{center}
\includegraphics[width=9cm]{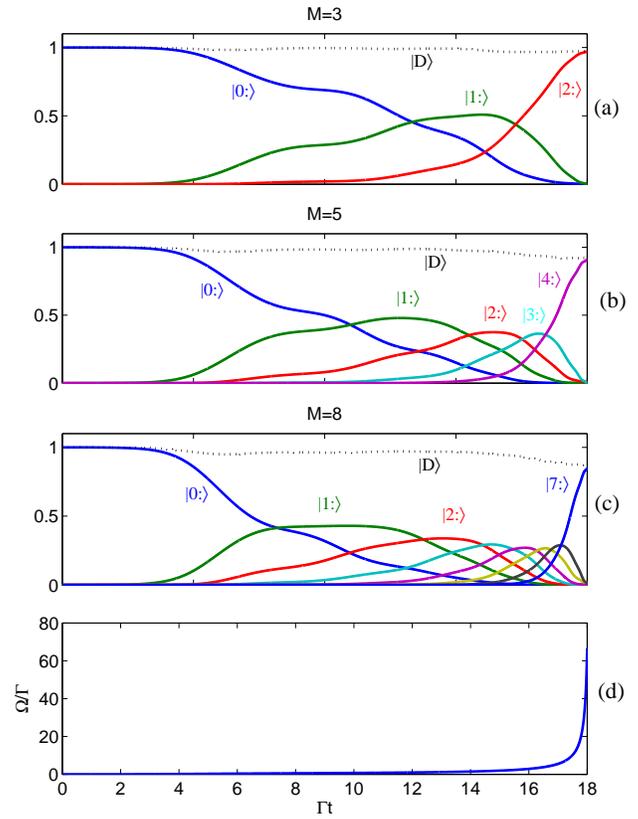}
\end{center}
\caption{(Color online) Time evolution of the system: For 3 (a), 5 (b),
and 8 (c) atoms. The dashed black curve is the population in dark-state.
The pumping field with a very steep shape has been chosen $\Omega (t)/\Gamma
=-3\left[ \tanh(4(\Gamma t-0.5))+1/(\Gamma t-2.01)\right]$.}
\label{fig:3}
\end{figure}

For our numerical calculations we have neglected the cavity loss and assumed $g=\Gamma$. The results are presented
in Fig.\ref{fig:2} and Fig.\ref{fig:3}.
From these numerical examples several conclusions can be drawn. First of
all one can see that the main part of population from initial state $|0:\rangle$
is adiabatically transferred into higher $|n:\rangle$ states when the amplitude
of the pumping field is increased. One can also see that the main part of
population stays in the dark state even for relatively large number of atoms
$M=8$. Better transfer can in principle be obtained with field changes
that satisfy the adiabaticity condition more strongly.

In Fig.3 numerical results obtained with pump field $%
\Omega (t)\sim \frac{1}{|t|}$ are presented. One might expect that these
type of pump fields are very inefficient due to large nonadiabatic
interactions arising in the region where the derivative of $\Omega (t)$ is
very large. However in our system the adiabaticity of the process increases
with $\Omega $ as can be concluded from eq.(\ref{adiabatic2}) which allows
us to apply pulses with very steep shapes. In this case the time that the
system spends in states with large value of $n_{ph}$ is smaller which strongly
reduces the cavity loss parameter (\ref{cavity_loss}).
For the results presented in Fig.3 the inequalities (\ref{adiabatic}) and
(\ref{adiabatic2}) are very well satisfied. The value of integral from (\ref%
{cavity_loss}) is $\Gamma\int_{0}^{T}n_{ph}\left( t\right) \mathrm{d}%
t\approx 5.3$ (a); $8.6$ (b) and $12.2$ (c). Thus proof of principle
experiments achieving non-trivial NOON states sizes can be realized if $\frac{g^2}{\kappa\Gamma}\approx10$,
although the influence of cavity decay cannot be completely disregarded.
Much better results can in principle be obtained with ultrahigh finesse
cavities with $\frac{g^{2}}{\kappa \Gamma } > 60$ \cite{Meschede-PRL-2009, Reichel-Nature-2007}.

Finally we would like to mention that the presented scheme works very much
in the same manner if initially all atoms of the system are in level
$\vert 1\rangle$ and the cavity contains a photon in arbitrary superposition
of modes $a$ and $b$ $\left(\left\vert \psi_{in}\right\rangle =\left( \alpha
\hat{a}^\dagger+\beta \hat{b}^\dagger\right)\vert vac\rangle\right)$.
It should also be noted that the presented scheme can be used to generate
mesoscopic entanglement in other systems such as trapped ions, provided
they are described by the Hamiltonian (\ref{Hamiltonian}).

To summarize we have discussed the evolution of two quantized modes
interacting with a medium of six level atoms
realizable as Zeeman sub-levels.
We have shown that this system has a dark eigenstate which is a mesoscopic
entangled state of photons and atoms. In the limit of strong coupling the
system can adiabatically be transferred to a dark state that can
be converted to a purely photonic or atomic NOON states by application
of simple local operations. We have also demonstrated that the efficiency
of NOON state generation increases exponentially with the single atom cooperativity
factor, thus allowing the preparation of large NOON states with moderate,
experimentally realized, cooperativity factors.

{\em Acknowledgements --} G.N. thanks Vladan Vuleti\'c for helpful discussion.
M.J.H. acknowledges funding from the DFG via the Emmy Noether project
HA 5593/1-1 and the CRC 631. This work was supported by the Alexander
von Humboldt Foundation, the EU STREP project HIP and the BMBF Verbundprojekt
QuoReP.

\end{document}